# Rigorous analysis of Casimir and van der Waals forces on a silicon nano-optomechanical device actuated by optical forces

Janderson R. Rodrigues[1], Andre Gusso[2], Felipe S. S. Rosa[3], and Vilson R. Almeida[1]

[1]*Instituto Tecnológico de Aeronáutica, Programa de Ciências e Tecnologias Espaciais*
*Praça Marechal Eduardo Gomes 50, CEP 12228-900, São José dos Campos, SP, Brazil*
[2]*Universidade Federal Fluminense, Departamento de Ciência Exatas*
*Avenida dos Trabalhadores 420, CEP 27255-125, Volta Redonda, RJ, Brazil*
[3]*Universidade Federal do Rio de Janeiro, Instituto de Física*
*Avenida Athos da Silveira Ramos 149, CEP 21941-972, Rio de Janeiro, RJ, Brazil*

In this article, we rigorously analyze the effects of the dispersion forces (Casimir and van der Waals forces) on a nano-optomechanical device based on a silicon waveguide and a silicon dioxide substrate, surrounded by air and driven by optical forces. The dispersion forces are calculated using a modified Lifshitz theory, in order to take into account the device thickness and material's dielectric permittivities, which are obtained from experimental optical data and validated by means of a rigorous 3D FDTD simulation. We also take into account the mechanical nonlinearity of the waveguide, which is caused by its large deflection relative to its thickness, due to the nanoscale device dimensions. The nonlinear mechanical analytical model is also validated using a 3D FEM simulation. Our results show that, under appropriate design conditions, it is possible to attain a no pull-in critical point due only to the optical force; therefore, in principle, it would be possible to control the device total deflection just by controlling the optical power. However, the dispersion forces usually impose a pull-in critical point to the device and establish a minimal initial gap between the waveguide and the substrate. Furthermore, we show that the geometric nonlinearity effect may be exploited in order to avoid or minimize the pull-in and, therefore, the device collapse.

## I. INTRODUCTION

Nano-optomechanical systems exploit light-matter interaction at the nanoscale. These devices are becoming very relevant to fundamental sciences, as well as to high-end technologies. In science, they have been achieving outstanding results both in the classical and the quantum regimes [1]. In technology, they may be applied in highly sensitive sensors, highly precise actuators, and active nanophotonic devices applications [2].

The mechanical motions of the majority of the nano-optomechanical devices are controlled by (transverse) optical (gradient) force, which arises from dipole moments induced in device's dielectric material, by the guided-light intensity gradient [3]. The nanometer dimensions allow strong interactions between the optical forces and the mechanical motion of such devices. The optical forces have been experimentally demonstrated in several dielectric nanodevices [4-10].

On the other hand, also due to the nanoscale separation between mechanical parts, the dispersion forces (Casimir and van der Waals forces) are the dominant interaction between uncharged materials. It may cause the collapse and adhesion between the movable parts which may occur during or after nanofabrication process, in a failure event known as stiction [11, 12]. The dispersion forces originate from quantum and thermal fluctuation of the electromagnetic field. The Lifshitz theory, based on the fluctuation-dissipation theorem, provides a unified description of both Casimir and van der Waals interactions between planar dielectric bodies [13]. In spite of the fact that dispersion forces have been recently experimentally demonstrated between dielectric nanomechanical devices [14-15], there is limited literature on the effects of the dispersion forces on nano-optomechanical devices [16-20].

In this work, we apply a modified Lifshitz theory to calculate the effects of the dispersion forces on a typical nano-optomechanical device, controlled by optical (gradient) forces, composed simply by a waveguide and a substrate; we also take into account the geometric nonlinearity during the waveguide deflection. Our rigorous and complete analysis leads to results that significantly modify or complement those found in the scientific literature on the subject [18].

## II. NANO-OPTOMECHANICAL DEVICE

This work considers a nano-optomechanical device formed by a silicon (Si) waveguide suspended over a silicon dioxide ($SiO_2$) substrate and surrounded by air, as illustrated in Fig. 1 and Fig. 2. This device can be fabricated, for instance, by using e-beam lithography, photolithography, and dry and wet etching processes on a SOI (Silicon-on-Insulator) platform [5]. It adopts the same structural dimensions used in [5, 18, 21, 22]. The Si waveguide has a length $L = 30$ μm, width $w = 500$ nm, and height $h = 110$ nm. The SiO$_2$ (optical) substrate has a height $H = 3$ μm and it is separated from the waveguide by an initial gap of $g_0 = 80$ nm. The underlying Si thick substrate of the SOI wafer does not play any role for this device, neither for optical nor for dispersion forces.

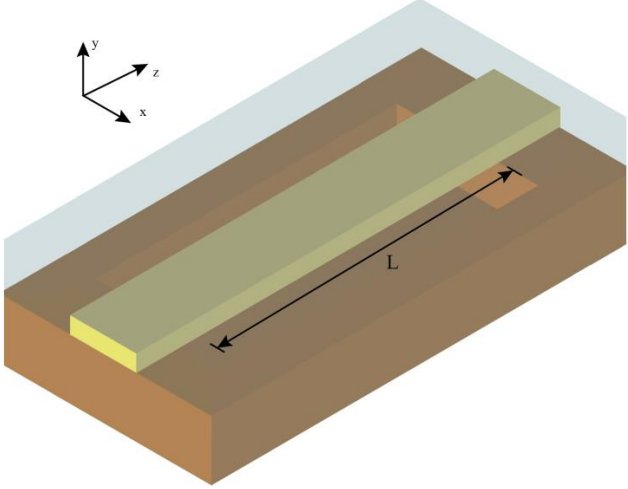

FIG. 1. Schematics of the nano-optomechanical device.

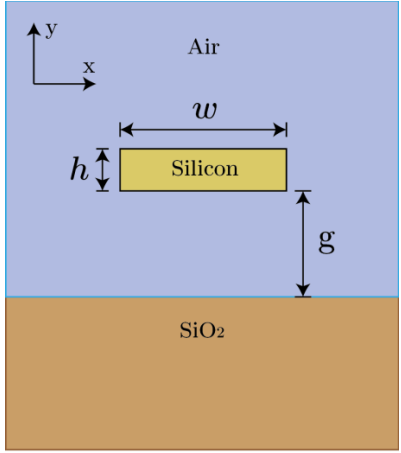

FIG. 2. Cross-section of the nano-optomechanical device.

## III. DISPERSION FORCES

The device we investigate involves a beam whose length and width are much larger than the nanowaveguide gap $g$. In such a case, it is reasonable to expect that the dispersive force could be approximated by that between two planar systems. In its turn, the beam thickness is comparable to $g$, both of the order of 100 nm, and we cannot immediately disregard the possible effect of the finite beam thickness on the dispersive force, and simply model the beam substrate interaction assuming two interacting semispaces with plane interfaces. The finite thickness effect must be investigated because significant thickness effects have been found previously for interacting dielectric planar layers [23][24]. Therefore, instead of the well-known expression derived originally by Lifshitz [13], for the dispersion forces between two semispaces separated by a finite gap, we resort to a modified expression found in the literature which allows us to account for the finite beam thickness.

A modified Lifshitz formula for the force per unit of area (pressure) on a system formed by two planar parallel layers, as illustrated in Fig. 3, composed by materials 2 and 4 with thicknesses $t_2$ and $t_4$, respectively, separated by a medium 3 with a gap distance g, and surrounded by two semi-infinite media 1 and 5, is given by [23] [24]:

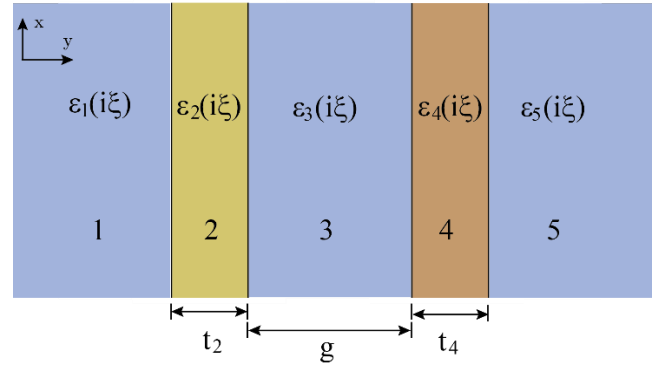

FIG. 3. Two planar parallel plates, composed by materials 2 and 4 separated by a medium 3 and surrounded by two semi-infinite media 1 and 5.

$$p_{dis}(g) = -\frac{k_B T}{\pi c^3} \sum_{l=0}^{\infty} \varepsilon_3^{3/2} \xi_l^3 \int_1^{\infty} p^2 \left\{ \frac{1}{\Delta} + \frac{1}{\bar{\Delta}} \right\} dp \quad (1)$$

where $\Delta$ and $\bar{\Delta}$ are the reflection coefficients given by:

$$\Delta = \Delta_{321}^{-1} \Delta_{345}^{-1} e^{2p\xi\sqrt{\varepsilon_3}g/c} - 1, \quad (2)$$

$$\bar{\Delta} = \bar{\Delta}_{321}^{-1} \bar{\Delta}_{345}^{-1} e^{2p\xi\sqrt{\varepsilon_3}g/c} - 1, \quad (3)$$

and the composed ones by,

$$\Delta_{321} = \frac{\Delta_{32} + \Delta_{21} e^{-2\xi\sqrt{\varepsilon_3}t_2 s_2/c}}{1 + \Delta_{32}\Delta_{21} e^{-2\xi\sqrt{\varepsilon_3}t_2 s_2/c}}, \quad (4)$$

$$\Delta_{345} = \frac{\Delta_{34} + \Delta_{45} e^{-2\xi\sqrt{\varepsilon_3}t_4 s_4/c}}{1 + \Delta_{34}\Delta_{45} e^{-2\xi\sqrt{\varepsilon_3}t_4 s_4/c}}, \quad (5)$$

with similar expressions for $\bar{\Delta}_{321}$ and $\bar{\Delta}_{345}$, respectively. In these equations $\hbar$ is the Planck constant divided by $2\pi$, $c$ is the speed of light in vacuum, $p$ is an auxiliary variable, $T$ is the absolute temperature and $k_B$ the Boltzmann constant. The $\xi_l = 2\pi k_B l/\hbar$ corresponds to the Matsubara frequencies. The reflection coefficients terms involve the product of the normalized difference of the materials' relative dielectric permittivities, $\varepsilon_k$,

$$\Delta_{jk} = \frac{(s_k - s_j)}{(s_k + s_j)}, \bar{\Delta}_{jk} = \frac{(s_k \epsilon_j - s_j \epsilon_k)}{(s_k \epsilon_j + s_j \epsilon_k)}, s_k = \sqrt{p^2 - 1 + \frac{\varepsilon_k}{\varepsilon_3}}, \quad (6)$$

where the subscripts $j$ and $k$ refer to the particular material of each layer, i.e., $j, k = 1, 2, 3, 4, 5$.

We note that by making $t_2 = t_4 \to \infty$ or, alternatively, $t_2 = t_4 = 0$ and forcing materials 1 and 5 becoming materials 2 and 4, respectively, in the previous equations, we recover the Lifshitz's original results for two semi-infinite plates, i.e. independent of the plates' thickness separated by medium 3 [13]. At small separations, of the order of a few nanometers, we achieve the non-retarded regime, in which case Eq. (1) reduces to the van der Waals force between two semi-spaces. On the other hand, at large distances, usually above several hundreds of nanometers for dielectrics and a few micrometers for metals we achieve the retarded regime, and Eq. (1) reduces to the Casimir force [25]. The van der Waals and Casimir forces between semi-spaces are characterized by distinct dependence on the gap, the first varying as $g^{-3}$ and the later as $g^{-4}$. For distances of a few tens of nanometers, as we consider in the system being investigated, there is no simple expression for the resulting dependence of the force on the gap $g$, and the Lifshitz theory must be used.

The permittivities entering in Eq. (1) are calculated along of the imaginary frequency $\xi$ at the imaginary axes $i\xi$, i.e. $\varepsilon_k = \varepsilon_k(i\xi)$. They can be evaluated using the Kramers-Kronig (KK) relation [26]:

$$\varepsilon_k(i\xi) = 1 + \frac{2}{\pi} \int_0^\infty \frac{\omega \varepsilon_k''(\omega)}{\omega^2 + \xi^2} d\omega, \quad (7)$$

where $\varepsilon_k''(\omega)$ is the imaginary part of the complex permittivities $\varepsilon_k(\omega) = \varepsilon_k'(\omega) + i\varepsilon_k''(\omega)$ as a function of the real frequency $\omega$. The function $\varepsilon_k(i\xi)$ is generally a real-valued monotonically decreasing function of its argument $\xi$, and it tends to the static permittivity $\varepsilon_k(i\xi) = \varepsilon_{k0}$ as $\xi \to 0$ and to unity $\varepsilon_k(i\xi) = 1$ as $\xi \to \infty$.

Following the procedures described in [27], we use optical data of the absorption spectra for monocrystalline Si ranging from far infrared to x-ray [28], which were interpolated and numerically integrated using Eq. (7). However, for $SiO_2$, we performed the Ninham-Parsegian (N-P) approximation to represent the infrared region [29],

$$\varepsilon(i\xi) = 1 + \sum_{u=1}^{n} \frac{C_u}{1 + (\xi/\omega_u)^2} \quad (8)$$

where $C_u$ is the absorption, $\omega_u$ is the resonance frequency, and $n$ is the total number of oscillators. This approximation is based on imaginary-axis representation of the frequency-dependent dielectric permittivity and on a harmonic (un)damped oscillator (disregarding the damping term) for lossless dielectrics. We considered the existence of three resonances in the infrared region ($n = 3$), namely $C_1 = 0.829$ and $\omega_1 = 0.867 \times 10^{14}$ rad s$^{-1}$, $C_2 = 0.095$ and $\omega_2 = 1.508 \times 10^{14}$ rad s$^{-1}$, and $C_3 = 0.798$ and $\omega_3 = 2.026 \times 10^{14}$ rad s$^{-1}$ [30]. For $SiO_2$ in higher frequencies, we apply the same procedure described for Si, using the spectral data available in [31, 32]. The dielectric permittivity of air was taken to be equal to that of vacuum $\varepsilon_{Vacuum}(i\xi) = \varepsilon_{Air}(i\xi) = 1$. In Fig. 4 we present the dielectric permittivities for silicon, silicon dioxide, and air.

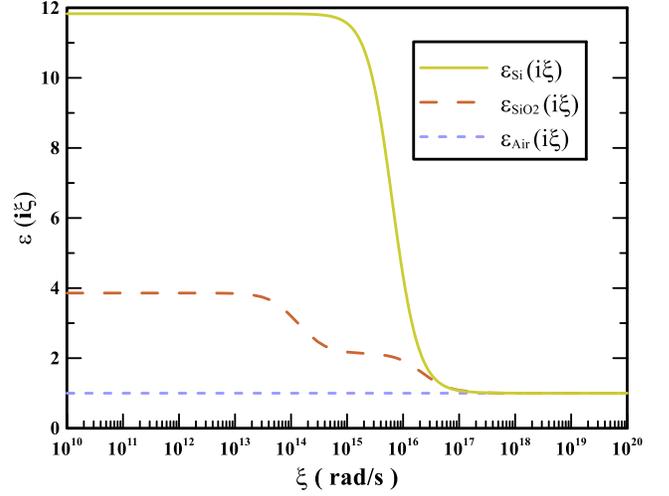

FIG. 4. (Color online) The dielectric permittivities $\varepsilon_k(i\xi)$ as function of the imaginary frequency $\xi$, for silicon (solid line), silicon dioxide (dashed line), and air (short dashed line).

The absolute dispersion pressure between a Si layer with a thickness of 110 nm and a SiO2 layer with 3 µm calculated using Eq. (1) at a temperature $T = 300$ K is presented in Fig. 6. In the range of interest, we have observed a very weak dependence of the force on the beam thickness. This is illustrated in Fig. 5 where we present the ratio $r$ between the force including the beam thickness (Eq. (1)) and that obtained for a Si and a SiO2 semispace, with the original Lifshitz equation [13]. The ratio is very close to 1 up to approximately 100 nm, revealing a negligible beam thickness effect in the system we investigate.

While the beam thickness effect can be accounted for by the modified Lifshitz theory for the planar model illustrated in Fig. 3, the theory does not account for effects that can result from the finite beam width. In order to verify the accuracy of this model, we compared its results with a more rigorous analysis using the finite-difference time-domain (FDTD) software Meep (*MIT Electromagnetic Equation Propagation*) [33-35]. We assumed that the 3D structure illustrated in Fig. 1 is *z*-invariant even under mechanical deflection, the validity of this assumption is discussed in the Section VI, what allows to reduce the computational time by means of the 2D model, as illustrated in Fig. 2, but indeed performing a 3D simulation [35]. The materials dielectrics permittivities were implemented on Meep extending the N-P approximation to the optical and ultraviolet regions for SiO$_2$ ($C_4 = 1.098$ and $\omega_4 = 203.4\times10^{14}$ rad) and using a single Lorentz resonance model for represent Si ($C_1 = 10.835$ and $\omega_1 = 6.6\times10^{15}$ rad). These values for Si have been obtained fitting the experimental results from Fig. 4 [35].

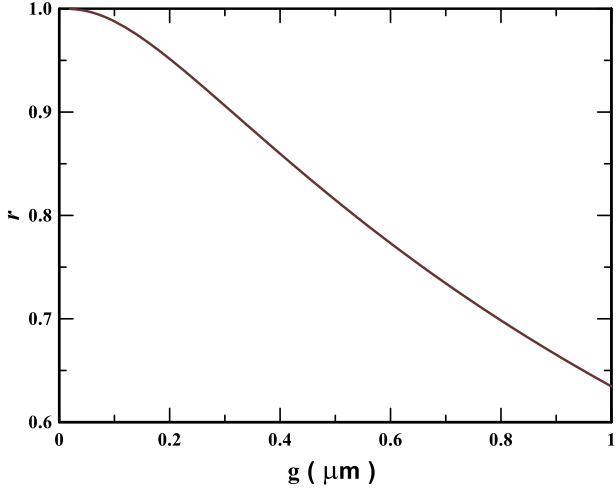

FIG. 5 (Color online) Finite beam thickness correction to the dispersion forces between layers made from Si and SiO2.

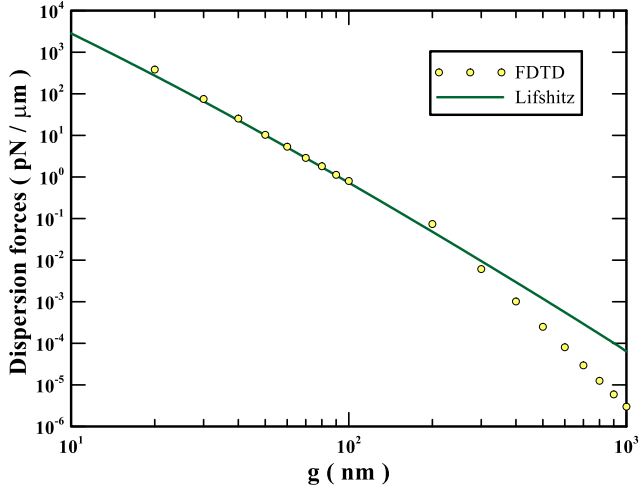

FIG. 6. (Color online) Dependence on the gap of the dispersion forces per unit length using the generalized Lifshitz theory (solid line) and FDTD simulations (circle).

In order to reduce the numerical error, first we performed the simulation only with the Si waveguide, then only with the SiO$_2$ substrate and, finally we subtracted these values of the simulation result of the whole structure (the waveguide and the substrate) [35]. The resulting dispersion forces per length unit ($F_{dis}/L = p_{dis}.w$) are presented in Fig. 6 showing an excellent agreement between the Lifshitz theory and the FDTD simulation for gap distances lower than the nanowaveguide' width (g < w). However, as the gap starts approximating the waveguide width, the boundary effects become important, leading the two models to diverge. However, since we are using g ≤ 80 nm and w = 500 nm, the Lifshitz model perfectly applies to the simulated nanodevice.

## IV. OPTICAL FORCE

The optical force between waveguides can be calculated either by means of the Maxwell Stress Tensor (MST) [3, 36-39] or, alternatively, in a linear and lossless media, directly from the device optical dispersion relation [3]:

$$F_{opt}(g) = -\frac{1}{\omega}\frac{d\omega}{dg}\bigg|_{\vec{k}} U(g), \quad (9)$$

where $U$ represents the total electromagnetic energy traveling through the waveguides in a given guided optical mode, $\omega$ is the angular optical frequency, and $\vec{k}$ is the associated guided wave vector; the derivative in Eq. (9) is taken at fixed wave vector. The total energy and the optical power $P$ are related by

$$U(g,P) = PL\frac{n_g(g)}{c}, \quad (10)$$

where $n_g$ is the group index. By substituting Eq. (10) into Eq. (9), we have

$$F_{opt}(g,P) = -\frac{1}{c}\frac{n_g(g)}{\omega}\frac{d\omega}{dg}\bigg|_{\vec{k}} PL. \quad (11)$$

However, for practical reasons, it is convenient to represent the optical force as a function of effective refractive index, $n_{eff}$ [37-39],

$$F_{opt}(g,P) = \frac{1}{c}\frac{dn_{eff}(g)}{dg}\bigg|_{\omega} PL. \quad (12)$$

Therefore, we have optical force as a function of effective index at a fixed optical frequency, i.e., at a fixed wavelength $\lambda$, since $\omega = 2\pi c/\lambda$. Notice that the optical force is attractive for $dn_{eff}/dg < 0$ and repulsive when $dn_{eff}/dg > 0$; besides that, the strongest dispersive dependence of $n_{eff}$ on $g$ will lead to larger optical force.

We adopted the Si refractive index $n_{Si} = 3.48$, the SiO$_2$ refractive index $n_{SiO2} = 1.44$, and the air refractive index $n_{Air} = 1.0$, at wavelength $\lambda = 1.55$ μm. The dispersion diagram of the structure, $n_{eff}$, as a function of the gap, $g$, for the structure presented in Fig. (2) was calculated for the fundamental (symmetric) quasi-TE eigenmode, using a commercial full-vectorial finite-difference mode solver (LUMERICAL MODE), for fixed wavelength of $\lambda = 1550$ nm [6]. Then, the normalized optical force was calculated using Eq. (12), and is shown in Fig. 7.

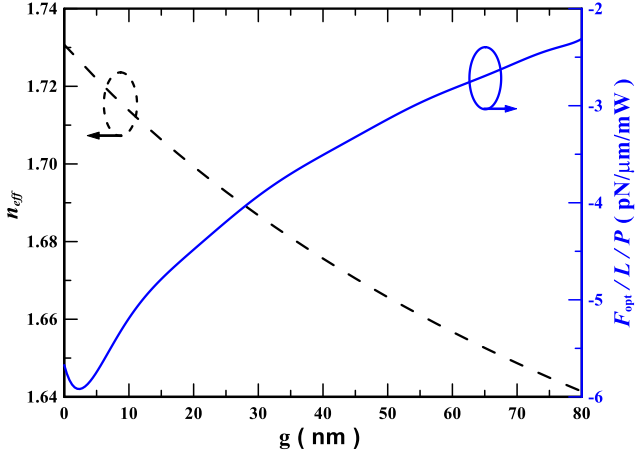

FIG. 7. (Color online) Effective refractive index (black line) and the normalized optical force (blue line) as a function of the gap.

For the symmetric modes, the optical force is always attractive (negative) for any value of gap. The normalized optical force obtained at null gap is 5.8 pN / μm / mW. Beyond that, the optical force values obtained in this case are much lower than those obtained for a Si slot waveguide [19] and, therefore, the optical power values to actuate the device must be much higher.

## V. MECHANICAL DEFLECTION

The nanowaveguide is fixed at its ends, forming a movable mechanical structure of a doubly-clamped beam. The deflection distributions $\delta_y(z)$, in the $y$ direction, along the waveguide length direction $z$, can be obtained through the Euler-Bernoulli beam equation for a distributed load with large deflection:

$$EI\frac{d^4\delta_y(z)}{dz^4} - N\frac{d^2\delta_y(z)}{dz^2} = \frac{F_y(g(z))}{L}, \quad (13)$$

where $E$ is the Young's Modulus, and $I$ is the area moment of inertia, in this case $I = wh^3/12$. For monocrystalline intrinsic silicon, the Young's Modulus varies with respect to crystallographic direction; here we considered the waveguide oriented in [100] direction in a (100) wafer, where $E = 130$ GPa.

The total applied load is given by $F_y = |F_{opt}| + |F_{dis}|$, with g replaced by $g(z) = g_0 - \delta_y(z)$. Due to the boundary conditions for the doubly-clamped beam, the deflection and slope at both ends are null, i.e., $\delta_y(0) = \delta'_y(0) = \delta_y(L) = \delta'_y(L) = 0$. In addition to this, the maximum deflection occurs at the middle of the beam, thus $\delta_{ymax} = \delta_y(L/2)$.

Due to the structure design, the maximum allowed deflection is given by $g_0$.

In the case of large deflections, the analysis includes a longitudinal axial force $N(\delta_y)$ that develops inside the beam. This axial force results in a nonlinear relation between the total load $F_y$ and its respective deflection $\delta_y$, also known as geometric nonlinearity. The maximum deflection for a uniformly distributed load can be found by solving simultaneously the next set of equations, [40]:

$$\frac{F_y(u)}{L} = 128\frac{EIh}{L^4}\sqrt{\frac{1}{3}}u^3(U)^{-\frac{1}{2}}, \quad (14)$$

$$\delta_y(u) = h\sqrt{\frac{1}{3}}(u - \tanh u)(U)^{-\frac{1}{2}}, \quad (15)$$

where

$$U(u) = \frac{2}{3} - \frac{3}{2}\frac{\coth^2 2u}{u} - \frac{1}{\sinh^2 2u} + \frac{1}{u^2}, \quad u = \frac{L}{4}\sqrt{\frac{N}{EI}} \quad (16)$$

where $u$ is a common variable and depends on the axial force $N$. The ratio between eq. (14) and (15) is the nonlinear mechanical stiffness and,

$$k_{NL}(u) = \frac{F_y(u)}{\delta_{y\max}(u)} = \frac{128EI}{L^3}\left(\frac{u^3}{u - \tanh u}\right). \quad (17)$$

Due to this nonlinearity, when the deflection increases, the spring constant of the beam becomes much larger than in the linear case. However, for small deflection we have no axial force $N = 0$, therefore $u = 0$, and this ratio tends to the linear mechanical stiffness;

$$k_L = \lim_{u \to 0}\frac{F_y(u)}{\delta_{y\max}(u)} = \frac{384EI}{L^3}. \quad (18)$$

The same result can be obtained by solving the Euler-Bernoulli equation for distributed load with small deflection limit by doing $N = 0$ directly in the Eq. (13).

The maximum deflection as a function of the linear and nonlinear elastic force, namely $F_{ela} = k_L\delta_{ymax}$ and $F_{ela}(u) = k_{NL}(u)\delta_{ymax}(u)$, is shown in Fig. 8. We also compare the analytical results obtained with Eqs. (14) and (15) for $N \neq 0$ and $N = 0$ with Finite Element Method (FEM) simulations in COMSOL Multiphysics software, using the full 3D device model, with and without the geometric nonlinearity option, respectively. The results show excellent agreement in both linear and nonlinear case. These results show that the small deflection theory of doubly-clamped beam under a distributed load is valid for deflections up to roughly a quarter of the beam thickness ($\delta_{ymax} = h/4$), which is the same limit obtained for a doubly-clamped beam under a concentrated load, but for much higher force values [40, 41]. The linear model for this value gives a percent error of 4% in relation to the nonlinear one; on the other hand, e.g., for deflections equal to the beam thickness $(\delta_{ymax} = h)$, it gives an error of 42%, for the material and dimensions utilized here. Hence, for a nanowaveguide with thickness $h$ of 110 nm subject to deflections up to a gap of 80 nm, the large deflection theory has to be used.

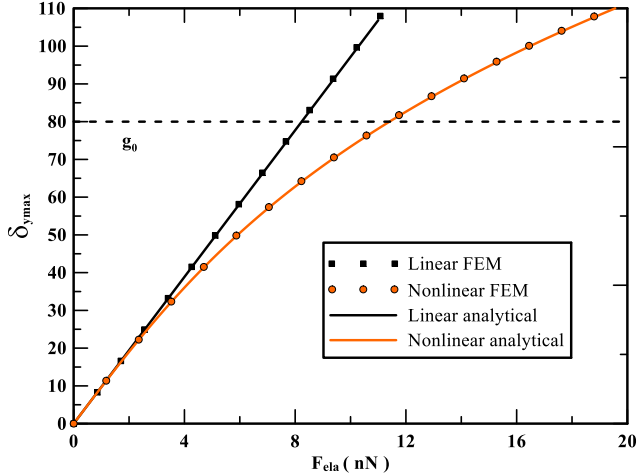

FIG. 8. (Color online) Deflection as a function of the elastic force for the analytical nonlinear (solid line) and linear (short dashed line) model and for FEM nonlinear (circle) linear (square) model.

## VI. PULL-IN INSTABILITY

The optical and dispersion forces interactions with the structure were obtained by solving Eq. (13) in the FEM package taking into account the load distribution variation along the beam length due to its own deflection. Here, in the dispersion forces, as well as in the optical force, we have assumed two perfectly parallel structures along the nanowaveguide length; however, when it starts to deform this condition is violated. Nevertheless, due to the doubly-clamped beam configuration, it has a small angle between the clamped part and the middle of the beam (slope), which implies a very small maximum deflection $\delta_{ymax}$ compared to the beam length $L$.

It is well known that the dispersion forces are not additive and it has strong geometric dependence [10]; however, in some conditions, it is possible to use the Proximity Force Approximation (PFA) method, where the curved surface is divided into infinitesimal parallel plates. The radius of curvature of the nanowaveguide, calculated using the maximum allowed deflection ($\delta_{ymax} = g_0$), is equals to $R = 1.41$ mm, which gives a curvature parameter of $\delta_{ymax}/R < 0.00006$; therefore, the PFA holds with an accuracy better than of 0.1% [42]. On the other hand, due to high refractive index contrast between silicon (core) and air (cladding), the bending and the propagation losses for this curvature radius and this nanowaveguide' length, respectively, are negligible; therefore, the mode effective index evolves adiabatically and its optical power remains constant during the propagation [3, 38].

Figure 9 shows a comparison between the absolute values of the dispersion forces calculated using Eq. (1), the nonlinear elastic force using Eqs. (14) and (15), and the optical force given by Eq. (12) acting upon the nanowaveguide, as a function its maximum deflection. The last force was calculated for the optical powers of 1, 10, and 100 mW. It is worth to notice that the dispersion forces magnitude is comparable with the optical force, for these typical values of optical powers, in this device. At zero deflection, the optical force is larger than the dispersion forces. As the deflection increases, this difference decreases, reaching the same value for a given optical power. After this point, the dispersion forces become dominant.

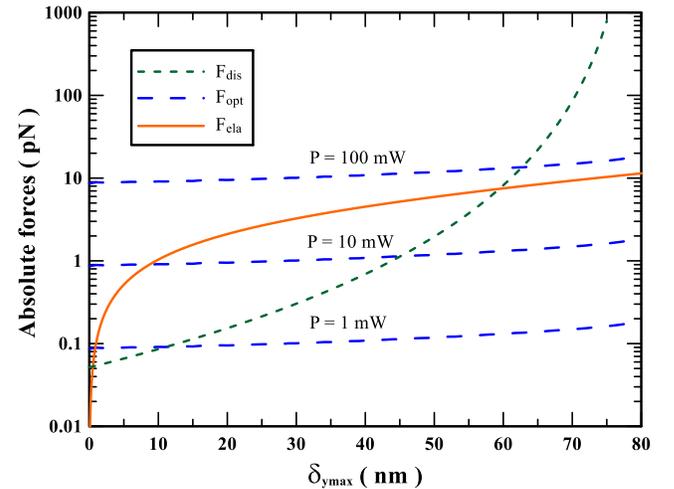

FIG. 9. (Color online) Dispersion forces (short dashed line), elastic force (solid line), and optical force (dashed line) for optical powers of 1, 10, and 100 mW, as a function of the maximum beam deflection.

The waveguide is initially (at null optical power) at static equilibrium, already under the effect of dispersion and elastic forces. Then, by applying an optical power, the structure undergoes a change in its equilibrium position, due to the attractive optical force. As the optical power increases, the combined optical and dispersion forces reach an equilibrium with the elastic force. This happens up to a certain maximum (critical) value of optical power ($P_C$), at which it attains the device's maximum (critical) deflection. Before this critical point, the system is stable, resulting in a net restoring force. However, beyond this point, the system becomes unstable, causing the collapse of the nanowaveguide with the substrate, due to the increasingly stronger attractive net force. As the optical power

(deflection) increases to the critical point, the pull-in initiates causing the waveguide to collapse, as seen in Fig. 10 and Fig. 11.

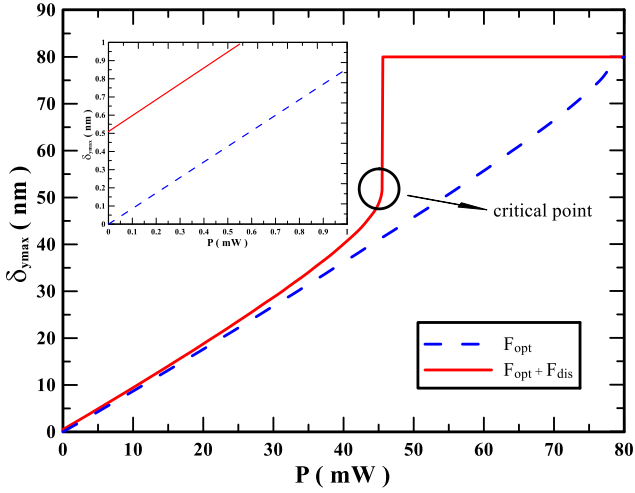

FIG. 10. (Color online) Maximum deflection as a function of the optical power of the optical force (dashed line) and the optical + dispersion forces (solid line). The inset shows the change in the initial equilibrium position.

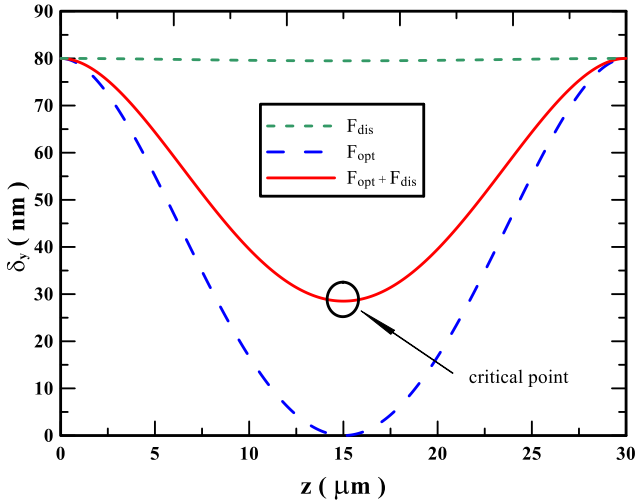

FIG. 11. (Color online) Deflection distribution along the waveguide under the influence of the dispersion forces (short dashed line), the optical force (dashed line) and the optical + dispersion forces (solid line).

Figures 9 and 10 show that, due to the nonlinear behavior of the elastic force, there is practically no pull-in caused only by the optical force; therefore, in principle, it would be possible to totally control the nanowaveguide deflection until its contact with the substrate solely adjusting the optical power. Just for comparison, if a linear deflection model were used instead, there would be a pull-in point at $P_C = 57.5$ mW ($\delta_{ymax} = 76.1$ nm) caused only by the optical force, which is prevented by the geometric nonlinearity. However, as seen in Fig. 10 the dispersion forces established a critical point for this specific structure at $P_C = 45.5$ mW ($\delta_{ymax} = 51.5$ nm) after what the collapse occurs. Using a linear model of the beam we would get instead $P_C = 38.5$ mW ($\delta_{ymax} = 48.1$ nm). Furthermore, even in the absence of optical power, dispersion forces cause changes in the initial equilibrium position of the structure, with $\delta_{ymax} = 0.51$ nm, as shown on the inset in Fig. 10. For smaller initial gap $g_0$, this discrepancy in the equilibrium position of the structure can generate a maximum deflection larger than the critical distance, causing its collapse. Therefore, since the dispersion forces scale with $g^{-3}$ for small distances, which represents a much stronger gap dependence than that of the nonlinear elastic restoring force, it always stablishes a critical point, as well as creates a fundamental limitation to the initial gap distance. The critical point could be suppressed by an appropriate inclusion of structural deflection limiters, such as tiny bumpers. The critical optical power as a function of the initial gap distance is presented in Fig. 12.

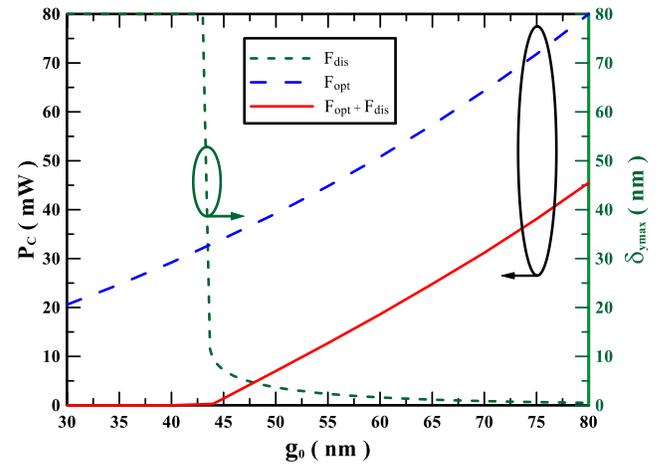

FIG. 12. (Color online) Critical Power and the maximum deflection as a function of the initial gap for the dispersion forces (short dashed line), the optical force (dashed line) and the optical + dispersion forces (solid line).

From Fig. 12, one can see that, as expected, the critical optical power decreases nonlinearly as the initial gap decreases. Besides that, the influence of the dispersion forces on the optical critical power increases as the gap decreases; at an initial gap of $g_0 = 43.7$ nm, the structure becomes unstable even in the absence of optical power, causing its collapse. Besides that, in the same manner that the geometric nonlinearity effect can avoid the pull-in caused only by the optical force, under appropriate design conditions, it may be exploited to counter act the dispersion forces effects on the collapse of such nano-optomechanical devices.

## VII. CONCLUSIONS

In this article, we have rigorously analyzed the effects of the Casimir and van der Waals forces (dispersion forces) on a realistically modeled nano-optomechanical device driven by optical force. The dispersion forces are calculated using a modified Lifshitz theory, which was validated by FDTD simulation. We showed that, due to the geometric nonlinearity, there is no pull-in critical point caused only by the optical force; however, the dispersion forces impose a critical power and establish a minimal initial gap between the waveguide and the substrate. Furthermore, the geometric nonlinearity may be exploited to avoid or minimize the device collapse. The results and conclusions presented here can be extended to other nano-optomechanical structures and materials.

## ACKNOWLEDGEMENTS

This work was supported in part by the Coordenação de Aperfeiçoamento de Pessoal de Nível Superior (CAPES) through a doctoral scholarship for J. R. Rodrigues and visiting researcher sponsorship for V. R. Almeida, and in part by the Conselho Nacional de Desenvolvimento Científico e Tecnológico (CNPq) under Grants 310855/2016-0 and 483116/2011-4. We are grateful to Alejandro Rodriguez at Princeton University for useful discussions about the Meep simulations.